\documentclass[a4paper,twocolumn,superscriptaddress,prl,noshowkeys]{revtex4}

\usepackage{amsmath}
\usepackage{graphicx}
\usepackage{color}
\usepackage[unicode=true,bookmarks=false,breaklinks=false,pdfborder={0 0 1},backref=false,colorlinks=false]{hyperref}
\usepackage{breakurl}

\usepackage{lettrine}

\setcitestyle{super}

\makeatletter

\newcommand{\warwick}{
Department of Physics, University of Warwick, Coventry CV47AL, UK
}

\newcommand{\lecce}{
Istituto Italiano di Tecnologia, IIT--Lecce, Via Barsanti, 73010 Lecce, Italy
}

\newcommand{\nnl}{
NANOTEC, Istituto di Nanotecnologia--CNR, Via Arnesano, 73100 Lecce, Italy
}

\newcommand{\madrid}{
Departamento de F\'{i}sica Te\'{o}rica de la Materia Condensada, UAM, Madrid 28049, Spain
}

\newcommand{\fede}{
Dipartimento di Fisica, Universit\`{a} Federico II di Napoli, 80126 Napoli, Italy
}

\newcommand{\lkb}{
Laboratoire Kastler Brossel, UPMC-Paris 6, ENS et CNRS, 75005 Paris, France
}

\newcommand{\ucl}{
Department of Physics and Astronomy, UCL, London WC1E6BT, UK
}

\begin{document}

\title{
{
\usefont{OT1}{cmss}{m}{n}
{\LARGE 
\textbf{Vortex and half-vortex dynamics in a spinor quantum fluid of interacting polaritons}
}
}
}


\author{Lorenzo Dominici}

\email{lorenzo.dominici@gmail.com}
\affiliation{\nnl}
\affiliation{\lecce}

\author{Galbadrakh Dagvadorj}
\affiliation{\warwick}

\author{Jonathan M. Fellows}
\email{j.fellows@warwick.ac.uk}
\affiliation{\warwick}

\author{Stefano Donati}
\affiliation{\nnl}
\affiliation{\lecce}

\author{Dario Ballarini}
\affiliation{\nnl}

\author{Milena De Giorgi}
\affiliation{\nnl}

\author{Francesca M. Marchetti}
\affiliation{\madrid}

\author{Bruno Piccirillo}
\affiliation{\fede}

\author{Lorenzo Marrucci}
\affiliation{\fede}

\author{Alberto Bramati}
\affiliation{\lkb}

\author{Giuseppe Gigli}
\affiliation{\nnl}

\author{Marzena H. Szyma\'{n}ska}
\affiliation{\ucl}

\author{Daniele Sanvitto}
\affiliation{\nnl}

{
  } 

\maketitle

{
\usefont{OT1}{cmss}{m}{n}
\textbf{Spinorial or multi-component Bose-Einstein condensates may sustain
  fractional quanta of circulation, vorticant topological excitations
  with half integer windings of phase and polarization. Matter-light
  quantum fluids, such as microcavity polaritons, represent a unique
  test bed for realising strongly interacting and out-of-equilibrium
  condensates. The direct access to the phase of their
    wavefunction enables us to pursue the quest of whether half
  vortices ---rather than full integer vortices--- are the fundamental
  topological excitations of a spinor polariton fluid. Here, we are
  able to directly generate by
  resonant pulsed excitations, a polariton fluid carrying  
  either the half or full vortex states as initial condition, 
  and to follow their
  coherent evolution using ultrafast holography. 
  Surprisingly we observe a rich phenomenology that shows a stable evolution 
  of a phase singularity in a single component as well as in the full vortex state, spiraling, splitting and branching of the initial cores under different regimes and the proliferation of many vortex anti-vortex pairs in self generated circular ripples. This allows us to devise the interplay of nonlinearity and sample disorder in shaping the fluid and driving the phase singularities dynamics.}
}\\



\usefont{OT1}{cmr}{m}{n}

\lettrine{\textbf{V}}{ortices} and topological excitations play a crucial role in our understanding of the universe,
recurring in the fields of subatomic particles, quantum fluids, condensed matter and nonlinear optics, 
being involved in fluid dynamics and phase transitions ranging up to the cosmologic scale~\cite{Zurek1985}. 
Spacetime could be analogue to a superfluid~\cite{Liberati2014} and elementary particles 
the excitations of a medium called the quantum vacuum~\cite{Volovik2003};
whetever this modern view will take strength or not,
phase singularities (i.e., vortices) of a quantum fluid (e.g., of a superfluid)
are point-like and quantized quasi-particles by excellence.
Here we make use of a specific experimental ``quantum interface'': 
polariton condensates~\cite{Byrnes2014}, which
are bosonic hybrid light-matter particles consisting of strongly
coupled excitons and photons. The $\pm 1$ spin components of the
excitons couple to different polarisation states of light making the
Bose-degenerate polariton gas a spinor condensate.
The realisation of
exciton polariton condensates in semiconductor microcavities
\cite{Kasprzak2006,Balili2007} has paved the way for a prolific series of studies
into quantum hydrodynamics in two-dimensional
systems~\cite{Dreismann2014,Roumpos2012,Amo2011,Pigeon2011,Sanvitto2010,Amo2009,Lagoudakis2008}.
Microcavity polaritons are particularly
advantageous systems for the study of topological excitations in interacting superfluids, thanks
to the stronger nonlinearities and peculiar dispersive and dissipative features, with respect to both atomic
condensates and nonlinear optics.

\begin{figure*}[htbp]
  \centering \includegraphics[width=0.65\linewidth]{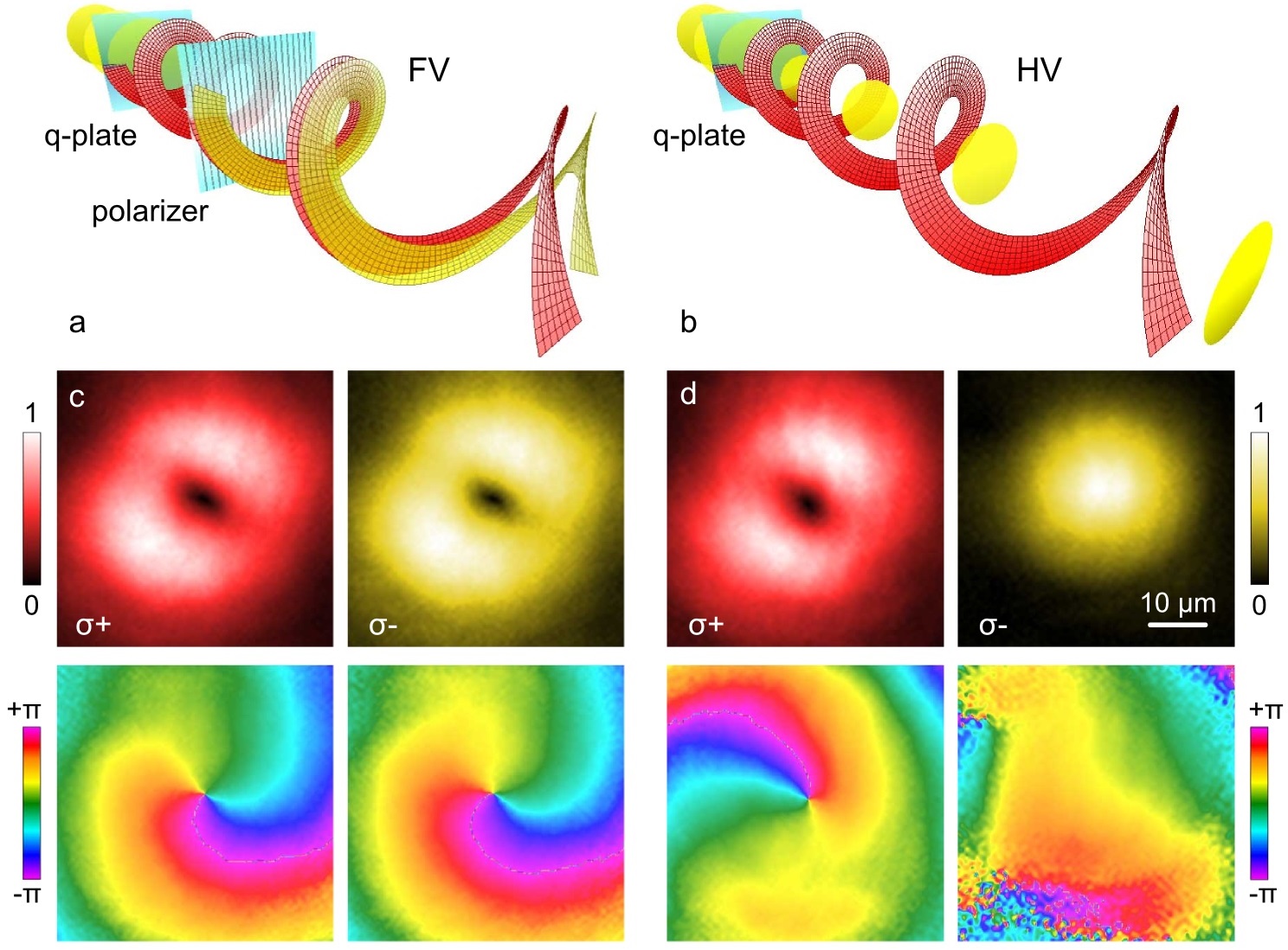} \linespread{1.1}
  \protect\protect\caption {Generation of optical and polariton FVs
    and HVs.  (a,b) Experimental scheme for creation of optical full-
    (a) and half-vortex (b) state via a q-plate. The disks and helics
    represent the isophase surfaces for Gaussian and vortex beams,
    respectively, in the radial regions of larger intensity.  Red and
    yellow colours refer to the $\sigma_{+}$ ($\sigma_{-}$) circular
    polarizations.  (c,d) Emission density of the polariton fluid at
    the time of initial generation and the corresponding phase maps.}
\label{fig:FIG1} 
\end{figure*}

For the equilibrium spinor polariton fluid, in which the drive and
decay processes are ignored, the lowest energy topological excitations
have been predicted to be ``half vortices''
(HV)~\cite{Toledo2014,Rubo2007}.  These carry a phase singularity in
only one circular polarisation, such that in the linear polarisation
basis they have a half-integer winding number for both the phase and
field-direction~\cite{Liu2015}. Such an excitation is complementary to a ``full
vortex'' (FV), which instead has a singularity in each circular
polarisation. Even in this simplified equilibrium scenario the
question of either HVs or FVs are dynamically stable has led to some
debate~\cite{Flayac2010a,Solano2010,Flayac2010b} due to the presence
of an inherent TE-TM splitting, which often arises in semiconductor
microcavities and couples HVs with opposite spin.~\cite{Toledo2014,Flayac2010a}.  
The issue is even more complicated
in a real polariton system, which is always subject to drive and
dissipation, and is intrinsically out of
equilibrium~\cite{Keeling2008}. Indeed, in the case of an incoherently
pumped polariton superfluid, in contrast to the equilibrium
predictions, it has been theoretically demonstrated~\cite{Borgh2010}
that both FV and HV are dynamically stable in the absence of a
symmetry breaking between the linear polarisation states, while in its
presence only full vortex states are seen to be stable. On the
experimental side, the recent work of Manni et al~\cite{Manni2012}
shows the splitting of a spontaneously formed linear polarised vortex
state (FV) into two circularly polarised vortices (HVs) under
non-resonant pulsed excitation. However, in this case, formation and
motion/pinning of these vortices are caused by strong inhomogeneities
and disorder in specific locations of the sample rather than 
by any fundamental process intrinsic to the fluid.  In general, the stability of
vortex states in polariton condensates remains an open issue of
fundamental importance, given that the nature of the elementary
excitations is likely to affect the macroscopic properties of the
system such as, for example, the conditions for the
Berezinsky-Kosterlitz-Thouless (BKT) transitions to the superfluid
state.  On the application side, polariton vortices have been proposed
also for ultra-sensitive gyroscopes~\cite{Franchetti2012} or
information processing~\cite{Sigurdsson2014}.\\

\textbf{Experimental system}\\
In this work we have been able for the first time to study the
dynamics of half and full vortices created into a polariton condensate
in a variety of initial conditions and in a controlled manner, taking
advantage of the versatility of the resonant pumping scheme. We take care to
generate the polariton vortex in a specific position on the sample
with sufficiently weak disorder that the biasing effects of sample
inhomogeneities can be screened out for a wide range of fluid densities.

In order to shape the phase profile of the incoming laser beam, we use 
a q-plate (Fig.~\ref{fig:FIG1}), a patterned liquid crystal
retarder recently developed to study laser windings and optical
vorticity~\cite{Marrucci2006,DAmbrosio2013,Cardano2013}. 
The q-plate allows us, through appropriate
optical and electrical tuning, to transform a Gaussian pulse into
either a FV or a HV, according to the simplified schemes shown in
Fig.~\ref{fig:FIG1}~a,b.  One advantage of a q-plate over using a
typical SLM (space light modulator) is evident in the fact that the
latter device works for a given linear polarization, and two SLM are
needed to create a HV.  The exciting pulse is sent resonant on the
microcavity sample to directly create a polariton fluid carrying
either a full or half vortex, as shown from the emission maps in
Fig.~\ref{fig:FIG1}~c,d.  Using a time-resolved digital
holography~\cite{Anton2012,Nardin2010} technique for the detection, we
measure both the instantaneous amplitude and phase of the polariton
condensate~\cite{Dominici2014} in all its polarization components.
Each phase singularity can be digitally tracked so as to record the evolution
of the resonantly created vortices after the initial pulse has gone
but before the population has decayed away. The lifetime of the 2D
polariton fluid in our microcavity sample
\cite{Dominici2013,Ballarini2013} kept at 10K is 10 ps, and we excite it by means
of a 80 MHz train of 4 ps laser pulses resonant with the lower polariton energy at 836 nm.\\

\begin{figure}[h]
\centering \includegraphics[width=7.8cm]{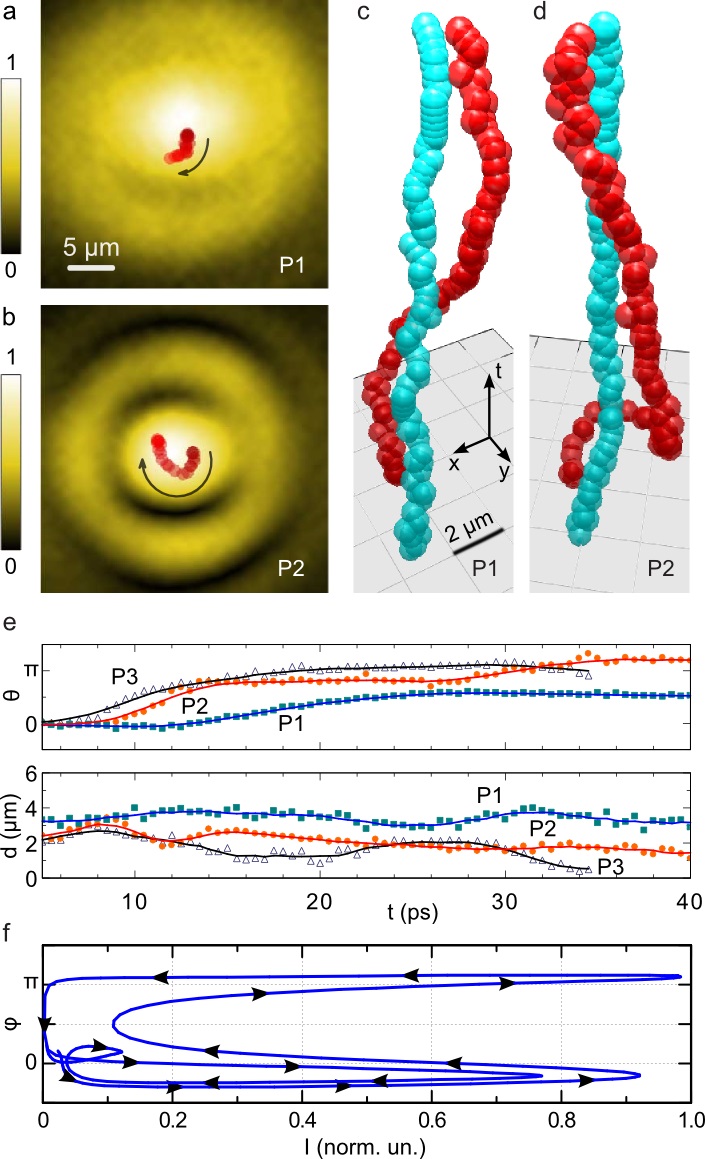} \linespread{1.1}
\protect\protect\caption{Evolution of the main singularity upon HV injection. 
The Gaussian map is shown 
together with the core trajectory in the opposite $\sigma$,
both 15 ps into the evolution, 
at the power of $P_{1}$=0.77 mW and $P_{2}$=1.8 mW in (a,b), respectively
(see also Supporting Movie SM1 for power $P_{1}$). 
The complete $(x,y,t)$ vortex trajectories (time range $\Delta t=5-40\text{ ps}$, 
step $\delta t=0.5\text{ ps}$) are shown in (c) and (d), 
with the blue spheres representing the Gaussian centroid 
and the red ones the phase singularity.
The angle $\theta$ and distance $d$
between the HV core and the opposite spin centroid 
are represented in (e) for 3 different powers.
Panel (f) is the phase-intensity plot along a vertical cut 
for $P_{2}$ and t=22 ps (arrows follow y),
higlighting a $\pi$-jump in the phase between adiacent maxima (i.e., when crossing the dark ring).}
\label{fig:FIG2} 
\end{figure}

\textbf{Dynamics of half and full vortex}\\
The creation of a HV is shown in Fig.~\ref{fig:FIG2} at different
pulse powers.  In panels (a,b) the trajectory of the primary vortex
($\Delta$t:5-15 ps, $\delta$t=0.5 ps) is superimposed to the amplitude map of
the opposite spin (taken at t=15 ps).  For both powers, the
singularity of the primary HV is seen moving along a circular
trajectory around the density maximum of the opposing Gaussian state,
keeping itself orbiting during few tenths of ps.  Such curves are
better depicted in Fig.~\ref{fig:FIG2} (c,d), which are the
$(x(t),y(t),t)$ trajectories ($\Delta$t:5-40 ps, $\delta$t=0.5 ps) relative to
cases (a,b), respectively, and in panel (e) reporting the angle
$\theta$ and distance $d$ between the primary HV core and the Gaussian
center of mass (see also Movie SM1).  

The orbital-like trajectories
suggests the presence of interactions between the vortex of $\sigma_+$
polaritons and the opposite $\sigma_-$ density.  Such dynamical
configuration resembles the metastable rotating vortex state,
predicted in~\cite{Ostrovskaya2012,Gautam2014}, supported by a
harmonic trap, although this effective potential is dynamically modified
by the intra-spin repulsive forces, e.g., by the deformation of the
initial Gaussian. Indeed the nonlinearities induce a breaking of radial symmetry, 
with the formation of circular ripples in the density.
The dark ripple shown in Fig.~\ref{fig:FIG2} (b) relative to the $\sigma_-$ Gaussian component,
presents a $\pi$-jump in the phase $\phi$, panel (f), 
which is a possible signature of a self-induced ring dark soliton (RDS),
considered its nonlinear drive. It is known that RDS are possible 
solutions of a 2D fluid with repulsive interactions~\cite{Rodrigues2014,Dominici2013,Manni2011}.
Yet, the displacement of the singularity (density
minimum) with respect to the centroid (opposite spin maximum), is
consistent with attractive inter-spin forces.  This is the first time that 
the manifestation of opposite spin interactions in polariton condensates 
is directly observed through their fluid dynamic effects.

\begin{figure}[h]
  \centering \includegraphics[width=8.4cm]{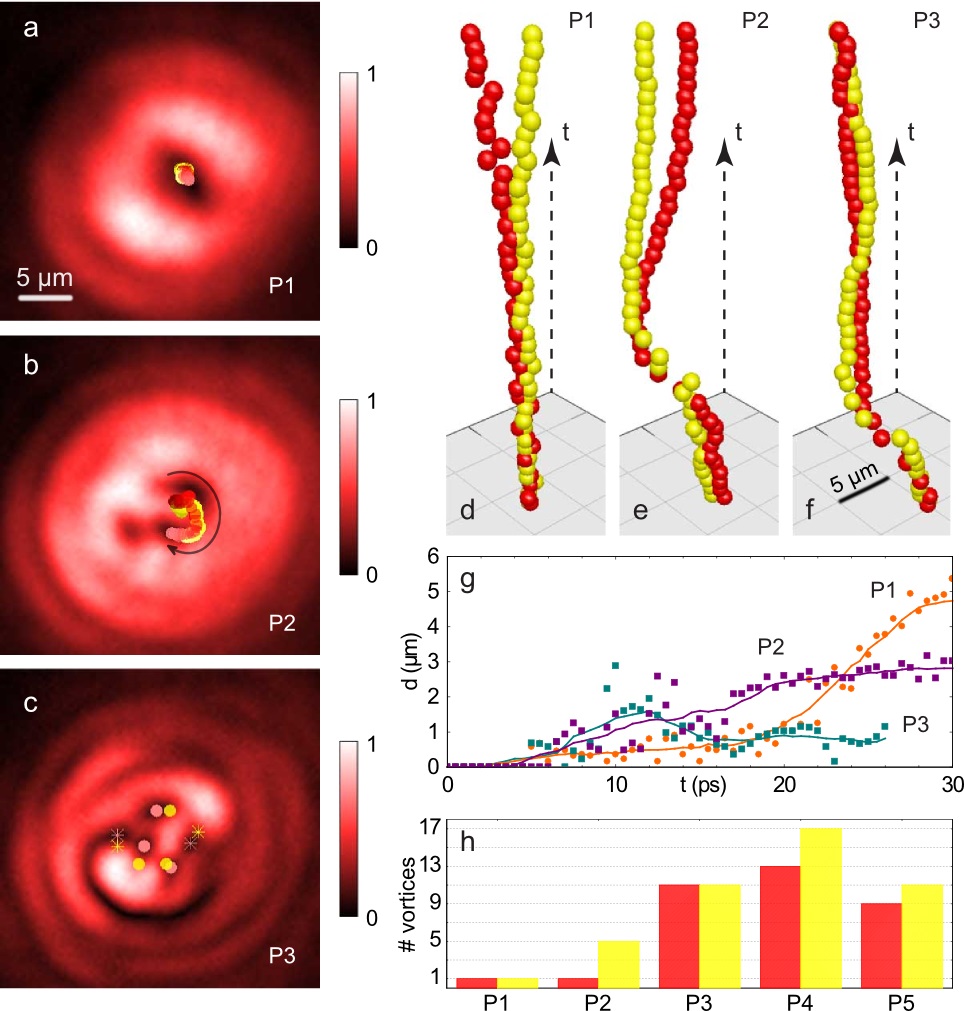} \linespread{1.1}
  \protect\protect\caption{Density maps and phase singularities upon
    resonant injection of FV states at different power regimes. (a-c)
    are the $\sigma_{+}$ density at $t=20\text{ ps}$ with superimposed
    phase singularities for both polarization, marked by symbols
    (circle for V, star for AV, colour for spin) (see Movies
    SM2--SM4). The trajectories of the primary vortices appear in (d-f)
    as 3D curves $(x,y,t)$ (time range $\Delta t=5-26\text{ ps}$, step $\delta t=0.5\text{ ps}$)
    (see Movie SM5 for $P_{1}$),
    and the evolution of the inter-core distance is resumed in
    (g). The final panel (h) shows the proliferation of secondary
    pairs (at $t=30\text{ ps}$) upon increasing pump power.
    The used laser powers are $P_{1-5}=0.17,~0.77,~1.8,~3.1~$and$~4.4~\text{mW}$,
    which correspond to an initial excitation of 
    $0.2, 1.0, 1.8, 2.2$ and $2.6\cdot10^6$ total polaritons, respectively.
    }
\label{fig:FIG3} 
\end{figure}

In Fig.~\ref{fig:FIG3} we show the generation of a vortex with
winding number $n=1$ for each circular polarisation ---i.e., a FV---
that can then be detected separately.  Panels (a-c) represent the
amplitude maps of one population ($\sigma_{+}$) at t= 20 ps with
superposition of the vortices positions (trajectories for (a,b),
instant positions for (c)) for three increasing pulse powers.  The
evolution of the primary singularities has been shown using 3D plots,
i.e., $(x(t),y(t),t)$ curves, in the panels (d-f) corresponding to
(a-c), respectively.  In the linear regime, at which the polariton
density is low (a,d, and Supporting Movie SM2), the opposite polarisation vortices evolve
jointly for the first few picoseconds once the pulse has gone.  As the
density starts to drop, the vortex cores show an increasing separation
in space, panel (g, orange), adopting independent trajectories.  This
suggests that the FV state is not intrinsically unstable, even though
it may undergo splitting supposedly driven by the sample disorder;
this is triggered when the density decreases below some critical value.

At larger polariton densities, Fig.~\ref{fig:FIG3}~(b,e) and Movie SM3, at which the
disorder is expected to be screened out, the twin singularities of the
injected FV move together while the fluid is reshaped under the drive of the nonlinear interactions and the
increase of radial flow.  Here, they also undergo a spiraling similar to the HV case.  
Interestingly, the twin cores appear to follow the same initial path, see also panel (g,
violet), hence indicating the lack of any intrinsic tendency of the FV
state to split.  This is confirmed by increasing the polariton density
further, Fig.~\ref{fig:FIG3}~(c,f) and (g, cyan), where the twin cores
remain together for even longer times. 
Any potential instability of a FV, and the consequent tendency to
split into two HV, is not observed here, 
differently from what observed in~\cite{Manni2012}, where the
splitting after non-resonant pumping was due to marked sample
inhomogeneities.  On the contrary our results show that at high
densities, for which the internal currents should prevail, there is a
strong inclination for the system to keep the full vortex state together.
However, note that the
increased density eventually causes circular density ripples, which
appear due to nonlinear radial currents and lead to the proliferation
of vortex-antivortex (V-AV) pairs in both polarisations.  In
particular, secondary vortices nucleate in the low density regions of
those circular ripples (Fig.~\ref{fig:FIG3}~c,h), which additionally disrupt
the original vortex core (see also Movie SM4).\\

\textbf{Theoretical Modeling}\\
In order to get a better understanding of the experimental vortex
  dynamics and interactions between the fundamental excitations, we have
  performed numerical simulations.  
The theoretical analysis performed by Rubo and collaborators in Ref.~\cite{Toledo2014} 
is based on the minimisation of the total energy 
for an equilibrium polariton condensate of infinite dimensions, i.e., 
where the density profile far from the vortex core is homogeneous. 
This analysis allows to establish a phase diagram for the stability/instability of different vortex excitations.
In contrast, here, we study the dynamics of finite size FV and HV states and their stability during
the dissipative and nonlinear evolution of interacting spinorial
 components, by dynamical simulations.  
We consider a generalised dissipative Gross-Pitaevskii equations 
for coupled two-component excitons $\phi_{\pm}(x,y,t)$ 
and microcavity photon $\psi_{\pm}(x,y,t)$ fields:
\begin{eqnarray}
  i\hbar\frac{\partial\phi_{\pm}}{\partial t} &
  =\left(-\frac{\hbar^{2}}{2m_{\phi}}\nabla^{2}-i\frac{\hbar}{\tau_{\phi}}\right)\phi_{\pm}
  +\frac{\hbar\Omega_{R}}{2}\psi_{\pm}\nonumber \\ 
  & + g|\phi_{\pm}|^{2}\phi_{\pm} + \alpha|\phi_{\mp}|^{2}\phi_{\pm}\label{eq:total}\\
  i\hbar\frac{\partial\psi_{\pm}}{\partial t} &
  =\left(-\frac{\hbar^{2}}{2m_{\psi}}\nabla^{2}-i\frac{\hbar}{\tau_{\psi}}\right)\psi_{\pm}
  +\frac{\hbar\Omega_{R}}{2}\phi_{\pm}\nonumber  
  \\ 
  & + D(x,y)\phi_{\pm} + \beta\left(\frac{\partial}{\partial x}\pm
    i\frac{\partial}{\partial y}\right)^{2}\psi_{\mp}+F_{\pm}\nonumber  
\end{eqnarray}
In order to reproduce the experimental conditions, we introduce a disorder term $D(x,y)$ for the photon field to match the inhomogeneities of the cavity mirror. The potential $D(x,y)$ is a Gaussian correlated potential with an amplitude of strength $50~\mu eV$ and a $1~\mu m$ correlation length.
 Since the effective mass of the excitons, $m_{\phi}$, is
4-5 orders of magnitude greater than that of the microcavity photons,
$m_{\psi}$, we may safely neglect the kinetic energy of the excitons.
The parameters in Eq.(1) are fixed so that to reproduce the experimental conditions, 
with an exciton and photon lifetimes of $\tau_{\phi}=1000~\text{ps}$ and
$\tau_{\psi}=5~\text{ps}$, respectively, a Rabi splitting
$\Omega_{R}=5.4~\text{meV}$ and the exciton-exciton interactions
strength $g=2~\mu\text{eV}\cdot\mu\text{m}^{2}$.  We take the
strength of the inter-spin exciton interaction to be an order of
magnitude weaker than the intra-spin interaction~\cite{Ferrier2011},
so that $\alpha=-0.1g$.  The coupling between different
  polarisations is given by the inter-spin interaction $\alpha$ and
by the TE-TM splitting term $\beta$.  Following Hivet \emph{et
  al}~\cite{Hivet2012}, we fix the ratio between the two effective
masses $m_{\psi}^{\text{TE}}/m_{\psi}^{\text{TM}}$ to $0.95$ in order
to have an intermediate TE-TM splitting
$\beta=\frac{\hbar^{2}}{4}(\frac{1}{m_{\psi}^{\text{TE}}}-\frac{1}{m_{\psi}^{\text{TM}}})=0.026\times\frac{\hbar^{2}}{2m_{\psi}}$.
The initial laser pulse is modelled as a pulsed Laguerre-Gauss
$F_{\pm}$:
\[
F_{\pm}(\mathbf{r})=f_{\pm}r^{|n_{\pm}|}e^{-\frac{1}{2}\frac{r^{2}}{\sigma_{\text{r}}^{2}}}e^{in_{\pm}\theta}e^{-\frac{1}{2}\frac{\left(t-t_{\text{0}}\right)^{2}}{\sigma_{t}^{2}}}e^{i(\mathbf{k}_{\text{p}}\cdot\mathbf{r}-\omega_{\text{p}}t)}
\]
where the winding number of the vortex component in the $\pm$ polarisation
is $n_{\pm}$. The strength, $f$, has been selected so as to replicate
the observed total photon output. The $\sigma_{\text{r}}$ and $\sigma_{t}$
parameters were chosen in order to have  
space width and time duration (FWHM) of
the pump $20~\mu\text{m}$ and
$4~\text{ps}$, respectively, in line with the experimental settings.
The pump is slowly switched on into the simulation, reaching
its maximum at $t_{\text{0}}=5.5~\text{ps}$ and cut out completely
after $5\sigma_{t}$ so as to avoid any unintended phase-locking.
We follow the dynamics of both full and half vortices shined resonantly 
with the lower polariton dispersion at $\mathbf{k}_{\text{p}}=0$ and $\omega_{\text{p}}=-1$.

\begin{figure}[h]
  \centering \includegraphics[width=7.2cm]{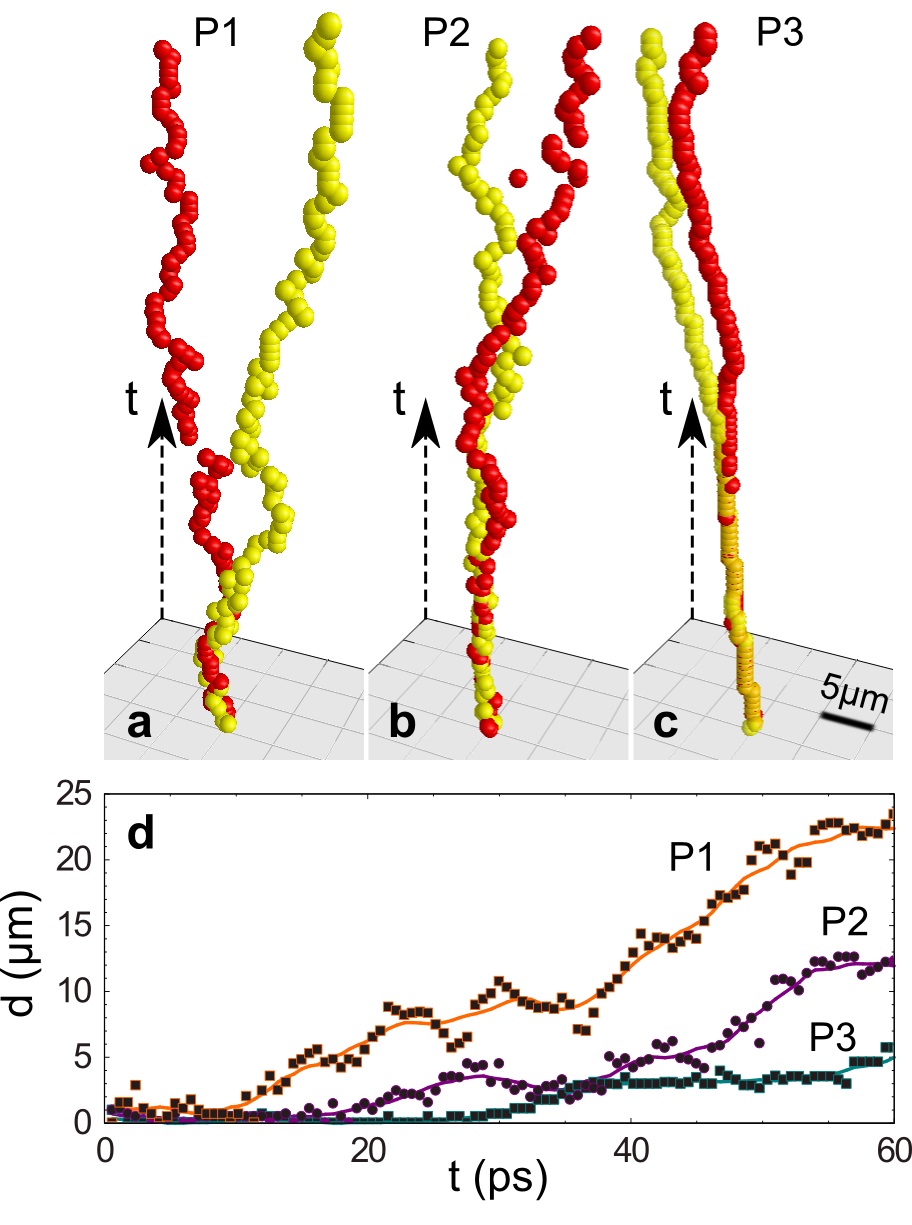} \linespread{1.1}
  \protect\protect\caption{Theoretical trajectories of primary singularities for
    FV state simulated at 3 increasing powers.  (a-c) are the 3D
    $(x,y,t)$ curves with $\delta t=0.4\text{ ps}$ step
    in a $\Delta t=0-60\text{ ps}$ span and the
    evolution of the inter-core distance is resumed in (d).}
\label{fig:FIG4} 
\end{figure}

Our simulations show that only in the presence of the disorder term
the imprinted vortex excitations undergo an erratic movement, both in the half
and full vortex configurations.  In agreement with the experiments 
the splitting of the FV is observed in the simulations only in the presence of disorder.
In Fig.~\ref{fig:FIG4} (a,b,c) we plot the trajectories for different
increasing powers $P_{1-3}$. 
The dissociation is seen at earlier times at low initial density, when the
sample disorder potential is expected to play a pivotal role. At
larger power the disorder and splitting are partially screened out, the
main charges move jointly for a longer time.  These results are
resumed in the panel (d) and are in a good qualitative agreement
with the experimental ones of Fig.~\ref{fig:FIG3}.
Simulations without disorder show that charges
are dynamically stable, immune to any internal splitting.  This holds
in our simulations even with artificially enlarged $\alpha$,
confirming that any dissociation is an external rather than an
intrinsic effect, at least during the polariton lifetime.  In
other terms, even though the thermodynamics would prefer HVs, based
upon energy minimization~\cite{Toledo2014}, the kinetics are too slow
to observe such effect in a real system.\\

\begin{figure}[h]
  \centering \includegraphics[width=8.4cm]{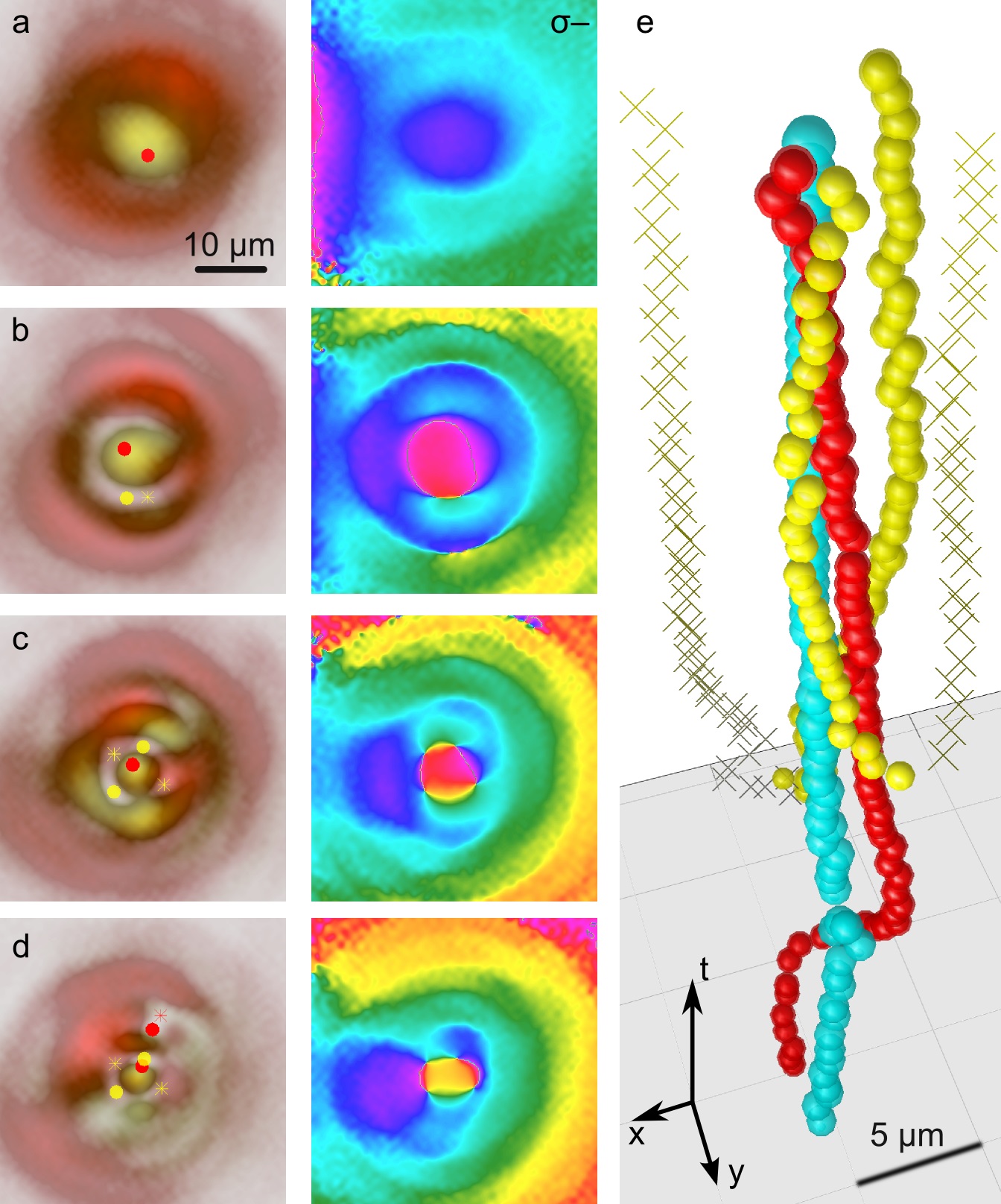} \linespread{1.1}
  \protect\protect\caption{Branching dynamics of a HV polariton
    condensate created at an intermediate power regime (1.8
    $\text{mW}$).  The four rows (a-d) show frames, taken at t = 8,
    16, 24 and 32 ps, with densities and vortices in the first column
    and associated phase maps for $\sigma_{-}$
    in the second column.  The initial
    condensate (a, orange due to overlap of red and yellow
    $\sigma_{\pm}$ intensity scale) undergoes the formation of
    concentric ripples (b-d, see also Movie SM6).  Spontaneous full
    V-AV formation with quadrupole symmetry for the initially Gaussian
    population is tracked and represented as $(x,y,t)$ vortex strings
    with $0.5\text{ ps}$ time step 
    in a $5-35\text{ ps}$ time span (e, see Movie SM7).}
\label{fig:FIG5} 
\end{figure}

\textbf{Branching and secondary vortices}\\
In the experiments, as already stated, at large densities both the HV and FV develop
concentric ripples, and this is causing
generation of secondary vortices.  For the HV, this effect is firstly
seen by increasing the power in the initially vortex-free Gaussian
component, where the same amount of total particles are concentrated
in a smaller area than in the vortex counterpart.  An exemplificative
case of this regime is shown in Fig.~\ref{fig:FIG5}, which reports in
the first column the overlapped density maps of the two populations
(red and yellow intensity scales) together with the vortices, and in
the second column the $\sigma_{-}$ phase
maps.  
The condensate evolves from the initial time (a), where only the
primary core of the HV is present, with the Gaussian developing more
marked ripples, generating a first V-AV couple (b) and then a second
one (c), which take positions in a 4-fold symmetric structure (see Movie SM6).
This effect is not driven by disorder. It is intrinsic and
  observed in a very large number of realizations and in different
  polarizations. Generation of secondary V-AV pairs is also seen in
  simulations, where the disorder term is removed (see Fig
  \ref{fig:FIG7}), confirming that this effect is not caused by the
  sample disorder.  The branching dynamics and its symmetry can be
clearly seen also in the 3D (xyt) trajectories of panel (e) (see also Movie SM7).  Only at
later time (d), when the density decreases substantially, also the
$\sigma_{+}$ component develops secondary pairs but in an external
region where the density drops locally.  It is worth noting that at
this later stage (d), the primary core of the HV, which was moving
around, is seen to merge with a secondary vortex of the opposite
polarization (but same winding), thus giving rise to the formation of
a FV.

\begin{figure}[h]
  \centering \includegraphics[width=8.4cm]{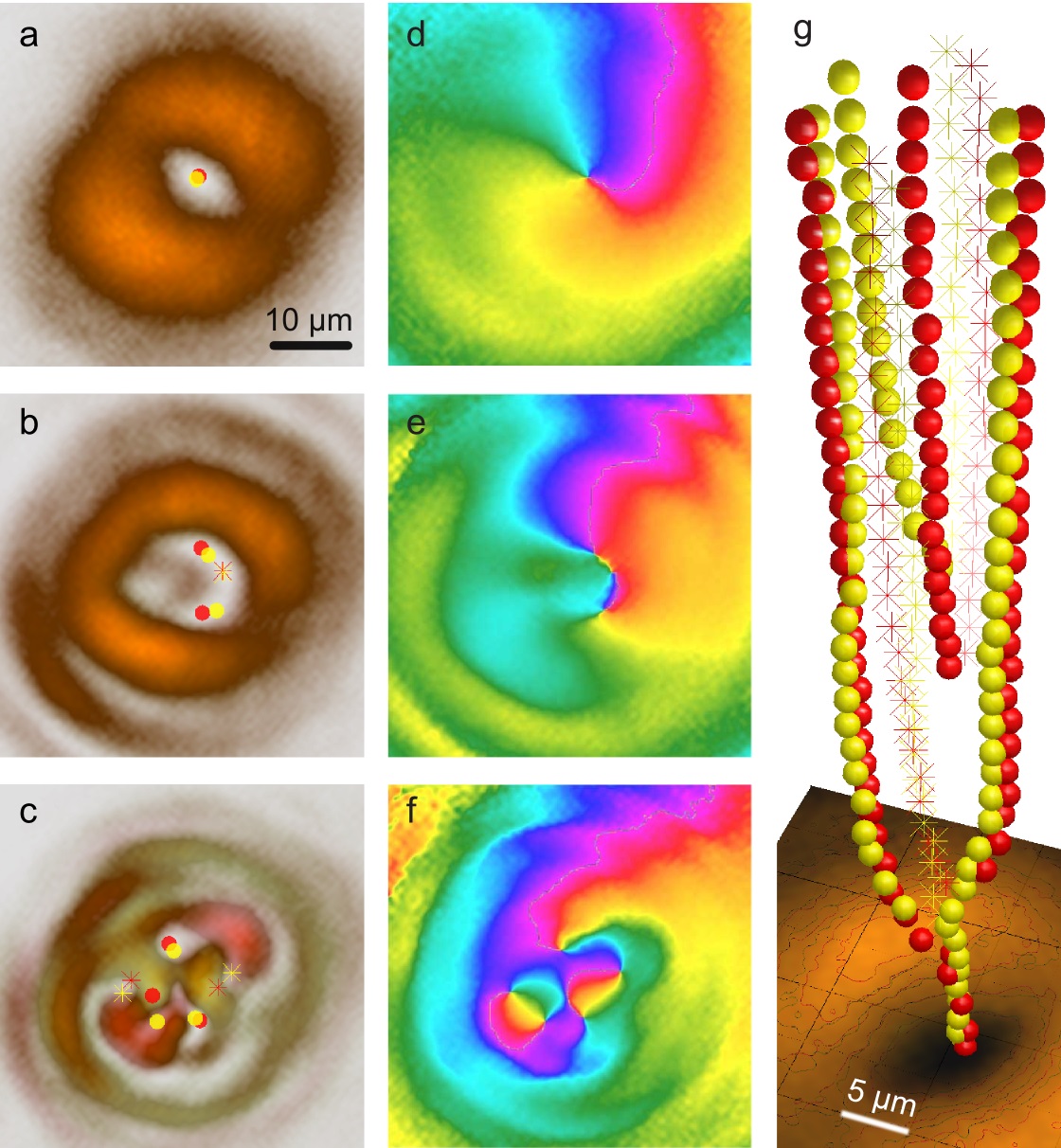} \linespread{1.1}
  \protect\protect\caption{Branching dynamics of a FV polariton
    condensate created at an intermediate power regime (1.8
    $\text{mW}$).  (a-c) are density frames and vortices taken at t =
    8, 12 and 24 ps, respectively, while (d-f) are the corresponding
    phase maps for just one polarisation ($\sigma_{-}$).  The initial
    condensate (a, orange due to overlap of red and yellow
    $\sigma_{\pm}$) develops concentric ripples (b-c, see also Movie
    SM4).  Spontaneous full V-AV formation is tracked as $(x,y,t)$
    vortex branches with time step of 
    $\delta t=0.5\text{ ps}$
    and $\Delta t=6-24\text{ ps}$ range
    in (g, see Movie
    SM8), for both the populations.  Each secondary HV stay close to
    its spin counterpart until quite late into the dynamics.}
\label{fig:FIG6} 
\end{figure}

The generation of secondary vortices is seen also in case of the FV,
as shown in Fig.~\ref{fig:FIG6}, at $P=1.8~\text{mW}$.  The panels
(a-c) represent the joint population and vortices at different time
frames, while the corresponding phase maps (d-f) are reported only for
one polarization.  We observe that while the primary FV (a)
rotates, it undergoes a displacement a moment before the creation of
the first V-AV pair, which is followed by a second one, (b) and
(c), respectively.  The two secondary V-AV pairs are created in
succession, and jointly between the two $\sigma$ states: in other
terms, the secondary topological charges are created as full vortex and
anti-vortex.  The $\sigma_{+}$ and $\sigma_{-}$ cores of the primary and the
first secondary FVs move together in a FV configuration for quite a
long time.  The branching and its partial symmetry, can be seen also
in the phase maps (d-f), and in the branch structure of
Fig.~\ref{fig:FIG6}(g) (see Movie SM8), with the $xyt$ trajectories of the vortices.
At later times the central region, initially dark, is partially filled
with fluid and some degree of asymmetry is present between the two
polariton distributions.  We found that at different densities,
localized transient structures with 3, 4 or 6-fold symmetries may
arise too (see also~\cite{Keeling2008}).

\begin{figure}[h]
  \centering \includegraphics[width=7.4cm]{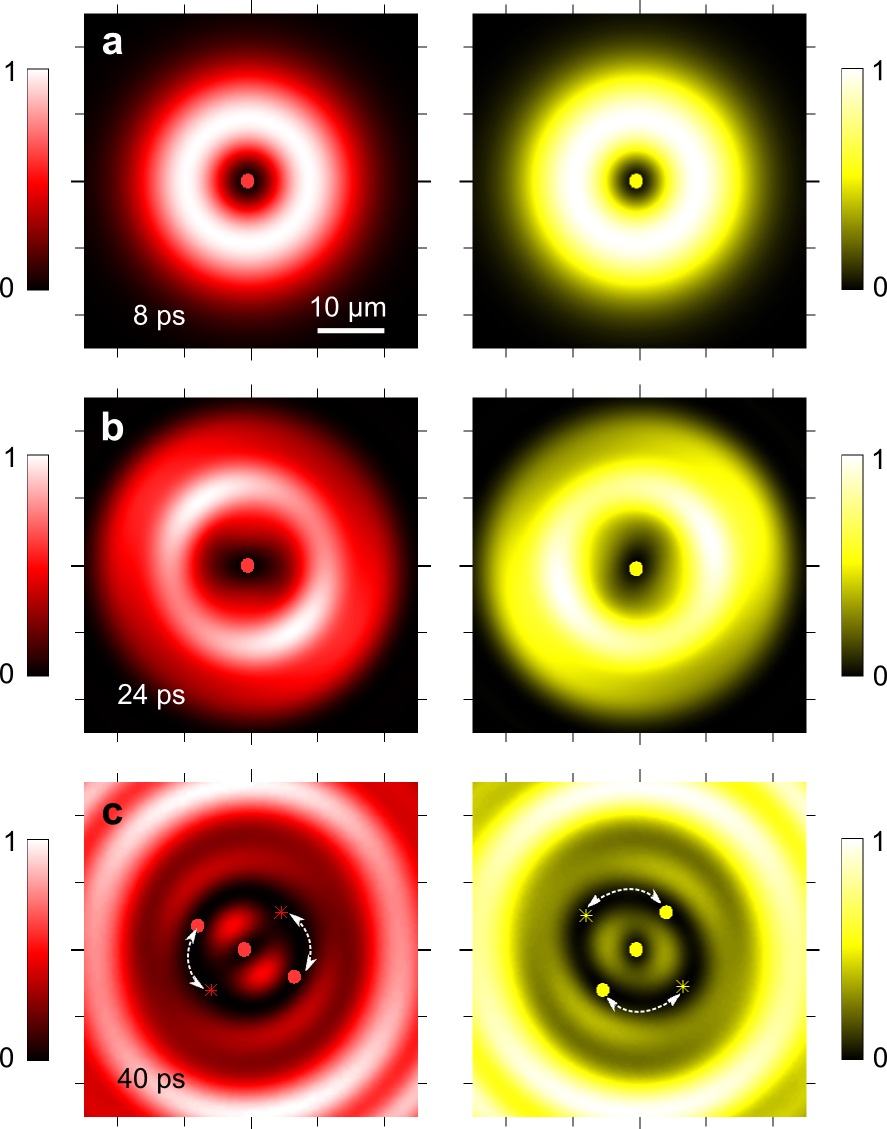} \linespread{1.1}
  \protect\protect\caption{Theoretical density maps and phase
    singularities in the case of FV without the disorder
      potential.  (a-c) Each row corresponds to a different time, (a)
    $t=8\text{ ps}$, (b) $24\text{ ps}$ and (c) $40\text{ ps}$.  Left
    and right columns represent the $\sigma_{+}$ and $\sigma_{-}$
    density, respectively, with superimposed their phase
    singularities, marked by symbols (circle for V, star for AV,
    colour for spin, see Movie SM9).}
\label{fig:FIG7} 
\end{figure}

In the simulations we see the emergence of density ripples (radial
symmetry breaking), as observed in the experiment, above certain
density (pump power) threshold with or without the disorder.  It is in
the very bottom of these ripples, where the density is almost zero,
that spontaneous V-AV pairs nucleate.  Figure \ref{fig:FIG7} shows the theoretical
evolution of the density maps for the two components of a FV, on each
column, respectively (see also Movie SM9). The main difference compared with experiments is that
here the secondary couples are generated in different positions for the two polarisations.
Yet, they keep rotating along a direction depending on their winding,
and not on their spins.  We have reasons
to believe that the direction of circulation could be associated to
the winding sign and the direction of the fluid reshaping (i.e.,
contracting or expanding), but the study of such aspect is well beyond the scope
of the present work.\\

\vspace{-0.5cm}

\textbf{Conclusions}\\
To conclude, we have investigated the dynamics and branching of half
and full vortices resonantly injected in an out-of-equilibrium
polariton quantum fluid.  The dynamics of these topological defects
is ruled by the interplay between the non-linearity and the disorder landscape. 
Our main conclusion is that, surprisingly, both FV and HV states
are intrinsically dynamically stable, i.e., the topological charges in the
two spin components do not split because of intrinsic energy
considerations during the lifetime of the polaritons, nor the
singularity of a half-vortex is seen to attract an opposite spin
counterpart. The splitting effects, we observe, can be attributed to
the fact that at low density (long time) the fluid streamlines are
affected more by the sample landscape, with disorder guiding the
displacement of the vortices, and eventually separating the cores when
a symmetry breaking term such as anisotropic or TE-TM splitting is at
action.  At intermediate density regimes, when sample inhomogeneities
are screened out and nonlinear turbulence is moderate, the charges
stay together for longer times.  It is at even larger densities, when
the main charges stay together up to tens of ps, that they are also
seen to move in a marked precessing trajectory, both for the HV and FV states.  
Here, the nonlinearities drive radial flows with the reshaping of the fluid
into circular ripples of alternating high and low density regions,
where secondary vortices nucleate.  This nucleation is systematic and
distinct from the proliferation of vortices at very low densities,
which are pinned by disorder,
as demonstrated by the theoretical simulations performed in an homogeneous landscape
 ---the secondary charges nucleate in
pairs of opposite winding in each of the two spin populations, and
their evolution is seen as quasi-ordered branching of 3D (2D+t) singularity trees.  
Our observations suggest that quantum
phase-singularities might be seen as an analogue of fundamental
particles, whose features can span from quantized events such as pair creation and
recombination to vortex strings.  Moreover, with
both topological states seemingly stable during the typical polariton
lifetimes, an interesting question left to be addressed is which
excitations are relevant for the Kosterlitz-Thouless-type transition
in these systems.\\

\textbf{Acknowledgments}\\
We acknowledge Giovanni Lerario for fruitful discussions, R.~Houdr\'{e} for the growth of the microcavity
sample and the project ERC POLAFLOW for
financial support. MHS acknowledges support from EPSRC (EP/I028900/2 and EP/K003623/2).
FMM acknowledges financial support from the Ministerio de Econom\'ia y
Competitividad (MINECO), projects No.~MAT2011-22997 and
No.~MAT2014-53119-C2-1-R.\\\\

\textbf{Supplemental Information}\\
Supporting Movies are available online at \href{https://drive.google.com/folderview?id=0B0QCllnLqdyBfmc2ai0yVF9fa2g2VnZodGUwemVkLThBb3BoOVRKRDJMS2dUdjlZdkRTQkU}{this URL}~.


\end{document}